\documentclass[a4paper,12pt]{article}
\usepackage{amssymb,amsmath}
\usepackage{epsfig,fancyhdr}

\usepackage{a4}
\usepackage{parskip}
\usepackage{color,soul,rotating}
\usepackage{amsthm,amsxtra,amscd,relsize,multirow}
\usepackage[all,2cell]{xy}\UseAllTwocells

\newcommand{\sect}[1]{\setcounter{equation}{0}\section{#1}}

\def\t{\tilde}

\def\cosh{\mathrm{cosh}}

\def\axs{AdS_5\times S^5}
\newcommand{\eq}[1]{\begin{equation} #1 \end{equation}}
\newcommand{\al}[1]{\begin{align} #1 \end{align}}

\begin{document}
\begin{titlepage}
\markright{\bf TUW--11--10}
\title{Quadratic corrections to three-point functions}

\author{D.~Arnaudov${}^{\star}$ and R.~C.~Rashkov${}^{\dagger,\star}$\thanks{e-mail:
rash@hep.itp.tuwien.ac.at.}
\ \\ \ \\
${}^{\star}$  Department of Physics, Sofia
University,\\
5 J. Bourchier Blvd, 1164 Sofia, Bulgaria
\ \\ \ \\
${}^{\dagger}$ Institute for Theoretical Physics, \\ Vienna
University of Technology,\\
Wiedner Hauptstr. 8-10, 1040 Vienna, Austria
}
\date{}
\end{titlepage}

\maketitle
\thispagestyle{fancy}

\begin{abstract}
Following the recent progress on the calculation of three-point correlators with two ``heavy'' (with large quantum numbers) and one ``light'' states at strong coupling, we compute the logarithmic divergent terms of leading bosonic quantum corrections to correlation functions with ``heavy'' operators corresponding to simple string solutions in $AdS_5\times S^5$. The ``light'' operator is chosen to be the dilaton. An important relation connecting the corrections to both the dimensions of ``heavy'' states, and the structure constants is recovered.
\end{abstract}

\sect{Introduction}

One of the most active fields of research in theoretical physics in recent years has been the correspondence between the large $N$ limit of gauge theories and string theory, and particularly the AdS/CFT correspondence~\cite{Maldacena}. Many impressive results from the duality between type IIB string theory on $AdS_5\times S^5$ and ${\cal N}=4$ super Yang-Mills theory~\cite{Maldacena,GKP,Witten} have been obtained, but much more lies beyond our knowledge. One of the problems that lack proper understanding is the calculation of three-point functions of string states (dual to operators with large quantum numbers in the gauge theory) at strong coupling ($\sqrt{\lambda}\gg1$). Although the problem remains unsolved in general, recently there has been significant progress in the semiclassical calculation of two-, three-, and four-point correlators with two ``heavy'' states \cite{Janik:2010gc}--\cite{Lee:2011}. Extending these studies, we consider the bosonic quadratic fluctuations\footnote{The logarithmic divergent terms of fermionic quadratic corrections are known to cancel exactly the bosonic ones so that conformal invariance is preserved. For this reason, we concentrate only on bosonic fluctuations.} of correlation functions of two ``heavy'' operators and the dilaton, utilizing the methods for calculation of three-point correlators suggested in \cite{Janik:2010gc,Costa:2010}.

The paper is organized as follows. To explain the method, in the next section we give a short review of and extend the procedure for computing
semiclassically three-point correlators with dilaton ``light'' operator in the case of $\axs$. Next, we proceed with the calculation of quadratic fluctuations of correlation functions for some simple solutions. We conclude with a brief discussion on the results.

\sect{Calculation of three-point correlators}

We consider $\axs$ background with Poincare coordinates $(z,x)$ in $AdS_5$, so that the boundary is a four-dimensional Minkowski space with coordinates $x$. As was shown in \cite{Costa:2010} the partition function assumes the form
\begin{equation}
\tilde{Z}(x_i,x_f,\Phi_0)\approx\int DX\,D\gamma\,D\Phi\,e^{i\left(S_P[X,\gamma,\Phi]+S_{SUGRA}[\Phi]\right)}\,.
\label{stringgenfun}
\end{equation}
We consider fluctuations around given string solution $\bar{X}^{\mu},\,\mu=0,\dots,9$. Examining the relevant geodesic equation $\ddot{\lambda}^{\mu}+\Gamma^{\mu}_{\nu\rho}\dot{\lambda}^{\nu}\dot{\lambda}^{\rho}=0$ with $\lambda^{\mu}(0)=\bar{X}^{\mu}$, we define $\xi^{\mu}=\dot{\lambda}^{\mu}(0)$. Up to quadratic fluctuations (second order in $\xi$) the partition function \eqref{stringgenfun} should be modified to
\eq{
\tilde{Z}(x_i,x_f,\Phi_0)\approx\int D\Phi\,D\xi\,e^{i(S_P[\bar{X},\bar{s},\Phi,\xi]+S_{SUGRA}[\Phi])}\,,
}
where the classical solution $\bar{X}$ to the equations of motion with suitable boundary conditions corresponds to an operator ${\cal O}_A$ with large quantum numbers in the dual gauge theory. We will confine ourselves to a solution which is point-particle in AdS
\eq{
z=z(\tau)=R/\cosh\,\kappa\tau\,,\qquad
x=x(\tau)=R\tanh\kappa\tau+x_0\,.
\label{xzparticle}
}
As was pointed out in \cite{Janik:2010gc}
\eq{
\label{kappabc}
\kappa\approx\frac{2}{s}\log{\frac{x_f}{\varepsilon}}\,,\qquad R\approx x_0\approx\frac{x_f}{2}\,,
}
where $\varepsilon$ is an ultraviolet regulator and $\kappa$ is defined through $t=\kappa\tau$. In addition, $\bar{s}$ is the saddle-point value of the modular parameter $s$ on the worldsheet cylinder, whose minimization of area gives the two-point function. $\Phi$ denotes the supergravity fields, one of which is the dilaton $\phi$. It sources the operator ${\cal D}_\phi\equiv{\cal L}$, which has scaling dimension $\Delta=4$ in the leading semiclassical approximation, near the boundary. The Lagrangian $\cal L$ of the ${\cal N}=4$ SYM theory generates a deformation of the 't Hooft coupling $\lambda$ \cite{Costa:2010}. The bosonic Polyakov action to leading order in the coupling $g=\frac{\sqrt{\lambda}}{4\pi}$ is \cite{Foerste}\footnote{We work in conformal gauge.}
\al{
S_P[\bar{X},\bar{s},\Phi,\xi]&=S_P^{(0)}[\bar{X},\bar{s},\Phi]+S_P^{(2)}[\bar{X},\bar{s},\Phi,\xi]\,,\\ \nonumber
S_P^{(0)}[\bar{X},\bar{s},\Phi]&=-g\!\int_{-\bar{s}/2}^{\bar{s}/2}\!\!d\tau\!\int\!d\sigma\,e^{\phi/2}\,\eta^{\alpha\beta}
\partial_\alpha\bar{X}^A\partial_\beta\bar{X}^Bg_{AB}\,,\\
S_P^{(2)}[\bar{X},\bar{s},\Phi,\xi]&=-g\!\int_{-\bar{s}/2}^{\bar{s}/2}\!\!d\tau\!\int\!d\sigma\,e^{\phi/2}\,\eta^{\alpha\beta}\!
\left(D_\alpha\xi^AD_\beta\xi^Bg_{AB}+\partial_\alpha\bar{X}^A\partial_\beta\bar{X}^B\xi^C\xi^DR_{ACBD}\right),
\nonumber
}
where $g_{AB}$ is the background metric, $D_\alpha\xi^A=\partial_\alpha\xi^A+\Gamma^A_{BC}\xi^B\partial_\alpha\bar{X}^C$, and the Riemann tensor is defined as
\eq{
{R^A}_{BCD}\equiv\partial_C\Gamma^A_{BD}-\partial_D\Gamma^A_{BC}+\Gamma^E_{BD}\Gamma^A_{EC}-\Gamma^E_{BC}\Gamma^A_{ED}\,.
}
We want to calculate the correction to the partition function
\eq{
\tilde{Z}^{(2)}=\int D\xi\,e^{iS_P^{(2)}[\bar{X},\bar{s},\Phi,\xi]}\,.
}
Since $S_P^{(2)}$ is quadratic in $\xi$, $\tilde{Z}^{(2)}$ is equal to the determinant of an operator ${\cal O}$. In order to find the determinant we consider the heat kernel technique \cite{Foerste}, which utilizes the powerful method of $\zeta$-function regularization. The correction to the partition function can be expressed as the formal sum ($t$ is an auxiliary parameter)
\eq{
\log\tilde{Z}^{(2)}=\frac12\int\frac{dt}{t}e^{-{\cal O}t}=\frac12\int_{\t\varepsilon}^\infty\frac{dt}{t}\sum_{n=-2}^\infty a_nt^{\frac{n}{2}-1},
}
where $\t\varepsilon$ is an ultraviolet cutoff with dimension of mass. We are only interested in the logarithmic divergent part of the quadratic corrections. Therefore we concentrate on $n=2$, and get
\eq{\label{Z2}
\tilde{Z}^{(2)}\sim\t\varepsilon^{-a_2/2}\,,\quad
a_2=a_2[\bar{X},\bar{s},\Phi]=-\frac{1}{4\pi}\int_{-\bar{s}/2}^{\bar{s}/2}d\tau\int d\sigma\, e^{\phi/2}\eta^{\alpha\beta}\partial_\alpha\bar{X}^A\partial_\beta\bar{X}^BR^C_{\phantom{C}ACB}\,,
}
where $a_2$ is the relevant Seeley coefficient.

Let us start with calculating the logarithmic divergent quadratic correction to the two-point function. The detailed analysis in \cite{Janik:2010gc} shows that there is a subtlety in obtaining the string propagator, so that the classical solution for the cylinder coincides with the classical state. Therefore we have to work with
\begin{equation}
\tilde{S}_P=S_P-\int_{-s/2}^{s/2}d\tau\int d\sigma\,\Pi^A\dot{X}_A\,,
\label{NewAction}
\end{equation}
which is minus the integral of the Hamiltonian. It was shown in \cite{Bak:2011} that, strictly speaking, we have to use the Routhian instead of minus the Hamiltonian, but they coincide for our considerations. To obtain the correction to the two-point function we should examine the quadratic fluctuation of $\tilde{S}_P$, i.e., we should calculate the corresponding $\tilde{Z}^{(2)}$. It can be shown straightforwardly that again one can use \eqref{Z2} just by substituting $\eta^{\alpha\beta}$ with $\delta^{\alpha\beta}$. Following \cite{Janik:2010gc,Costa:2010}, we get for the quantum correction of the two-point function (and scaling dimension) of two ``heavy'' operators
\al{\nonumber
\langle{\cal O}_A(0){\cal O}^*_A(x_f)\rangle&\sim\left(\frac{\varepsilon}{x_f}\right)^{2\Delta_A}\!\!=\tilde{Z}(0,x_f,\Phi_0=0)\\ \label{2point}
&=\int D\xi\,e^{i\tilde{S}_P[\bar{X},\bar{s},\Phi=0,\xi]}=\left(\frac{\varepsilon}{x_f}\right)^{2\Delta_A^{(0)}}\!\!\t\varepsilon^{-\tilde{a}_2/2}\,,\\
\Delta_A&=\Delta_A^{(0)}+\Delta_A^{(2)}\,,\qquad\Delta_A^{(2)}=-\frac{\tilde{a}_2\log\t\varepsilon}{4\log\frac{\varepsilon}{x_f}}\,,
\nonumber
}
where we have used \eqref{Z2}, and having defined the ``modified'' Seeley coefficient
\eq{
\tilde{a}_2=\tilde{a}_2[\bar{X},\bar{s},\Phi=0]=-\frac{1}{4\pi}\int_{-\bar{s}/2}^{\bar{s}/2}d\tau\int d\sigma\, \delta^{\alpha\beta}\partial_\alpha\bar{X}^A\partial_\beta\bar{X}^BR^C_{\phantom{C}ACB}\,.
\label{tildea2}
}
The three-point correlation function at strong coupling of two ``heavy'' operators and one dilaton can be obtained by functional differentiation of the partition function with respect to the dilaton field
\eq{
\langle{\cal O}_A(0){\cal O}_A^*(x_f){\cal D}_\phi(y)\rangle\approx\frac{I_\phi[\bar{X},\bar{s};y]}{x_f^{2\Delta_A}}\,.
}
With a slight abuse of notation, we get for the logarithmic divergent part
\al{
I_\phi[\bar{X},\bar{s};y]&=I_\phi^{(0)}[\bar{X},\bar{s};y]+I_\phi^{(2)}[\bar{X},\bar{s};y]\,,\\
I_\phi^{(0)}[\bar{X},\bar{s};y]&=i\int_{-\bar{s}/2}^{\bar{s}/2}d\tau\int d\sigma\left.
\frac{\delta S_{P}^{(0)}[\bar{X},\bar{s},\Phi]}{\delta\phi}\right|_{\Phi=0}K_\phi(\bar{X};y)\,,\\
I_\phi^{(2)}[\bar{X},\bar{s};y]&=-\frac{\log\t\varepsilon}{2}\int_{-\bar{s}/2}^{\bar{s}/2}d\tau\int d\sigma\left.
\frac{\delta a_2[\bar{X},\bar{s},\Phi]}{\delta\phi}\right|_{\Phi=0}K_\phi(\bar{X};y)\,,
\label{Iphi}
}
where the bulk-to-boundary propagator has the following form \cite{Freedman:1998}
\eq{
K_\phi(\bar{X};y)=K_{\phi}(z(\tau),x(\tau);y)=\frac{6}{\pi^2}\!\left(\frac{z(\tau)}{z^2(\tau)+(x(\tau)-y)^2}\right)^4\!\!.
}

Also, the following relation was discovered in \cite{Costa:2010} for the leading semiclassical approximation
\eq{
\langle{\cal O}_A(0){\cal O}_A^*(x_f){\cal L}(y)\rangle\approx-\frac{g^2}{2\pi^2}\frac{\partial\Delta_A}{\partial g^2}\frac{x_f^{4-2\Delta_A}}{y^4(x_f-y)^4}\,,
}
where the conserved charges are assumed constant. As we shall see below, it holds even for the quadratic corrections
\eq{
I_\phi^{(2)}[\bar{X},\bar{s};y]=-\frac{g^2}{2\pi^2}\frac{\partial\Delta_A^{(2)}}{\partial g^2}\frac{x_f^4}{y^4(x_f-y)^4}\,.
\label{aLAA}
}

\sect{Quantum corrections to three-point functions}

In this section we apply the methods described in the previous one to particular solutions.

\subsection{Circular rotating string}

Let us consider the case of circular rotating string with two equal spins in the sphere \cite{Frolov:2003qc}. First, we fix the notation by writing down the explicit form of the metric
\eq{
ds^2_{\axs}/R_{\rm str}^2=\frac{dz^2+dx^2}{z^2}+[d\gamma^2+\cos^2\gamma d\varphi_3^2+\sin^2\gamma(d\psi^2+\cos^2\psi d\varphi_1^2+\sin^2\psi d\varphi_2^2)]\,.
}
If we assume the point-particle solution in $AdS_5$ \eqref{xzparticle}, the Polyakov action in conformal gauge can be written as
\al{\nonumber
S_P[X,s,\Phi]&=g\int_{-s/2}^{s/2}d\tau\int d\sigma\,e^{\phi/2}\,\{\kappa^2+\dot{\gamma}^2-{\gamma'}^2+\cos^2\gamma(\dot{\varphi_3}^2-{\varphi'_3}^2)\\
&+\sin^2\gamma[\dot{\psi}^2-{\psi'}^2+\cos^2\psi(\dot{\varphi_1}^2-{\varphi'_1}^2)+\sin^2\psi(\dot{\varphi_2}^2-{\varphi'_2}^2)]\}\,.
}
The solution has the following form \cite{Frolov:2003qc}
\eq{
\gamma=\frac{\pi}{2}\,,\quad\psi=\sigma\,,\quad\varphi_1=\varphi_2=\omega\tau\,,\quad\varphi_3=0\,.
}
It is straightforward to find the conserved quantities and dispersion relation for this string configuration
\eq{
J\equiv J_1=J_2=2\pi g\omega\,,\qquad E=2\sqrt{J^2+4\pi^2g^2}=\Delta_A^{(0)}.
}
The dual operator is of the type ${\cal O}_A\sim{\rm Tr}(X^{J_1}Z^{J_2})$.

Let us apply now the procedure outlined briefly in the previous section. For the logarithmic divergent part of the quadratic correction to the Polyakov action \eqref{Z2} we obtain
\eq{
a_2[\bar{X},\bar{s},\Phi]=\frac{1}{\pi}\int_{-\bar{s}/2}^{\bar{s}/2}d\tau\int_0^{2\pi}d\sigma\,e^{\phi/2}(\kappa^2-\omega^2+1)\,.
}
The modified correction \eqref{tildea2} takes the form
\eq{
\tilde{a}_2[\bar{X},\bar{s},\Phi=0]=2\int_{-\bar{s}/2}^{\bar{s}/2}d\tau\,(\omega^2-\kappa^2+1)\,.
}
The saddle point with respect to the modular parameter $s$ is given by \cite{Costa:2010}
\eq{
\label{saddlecircular}
\bar{s}=\frac{2i}{\sqrt{1+\omega^2}}\log\frac{\varepsilon}{x_f}\,,
}
which along with \eqref{kappabc} implies the Virasoro constraint $\kappa=i\sqrt{1+\omega^2}$. Thus, the evaluation of $\tilde{Z}(0,x_f,\Phi_0=0)$ \eqref{2point} gives
\eq{
\langle{\cal O}_A(0){\cal O}^*_A(x_f)\rangle\sim\left(\frac{\varepsilon}{x_f}\right)^{2\Delta_A}
=\left(\frac{\varepsilon}{x_f}\right)^{2\Delta_A^{(0)}-4i\sqrt{1+\omega^2}\log\t\varepsilon},\qquad\Delta_A=\Delta_A^{(0)}+\Delta_A^{(2)},
}
which leads to
\eq{
\Delta_A^{(2)}=-\frac{i\sqrt{J^2+4\pi^2g^2}}{\pi g}\log\t\varepsilon\,.
}

Now we turn to the derivation of the fluctuation of the three-point function. We evaluate \eqref{Iphi} at the saddle point \eqref{saddlecircular}
\eq{
I_\phi^{(2)}[\bar{X},\bar{s};y]=\frac{3\omega^2\log\t\varepsilon}{\pi^2}\int_{-\bar{s}/2}^{\bar{s}/2}d\tau\left(\frac{z}{z^2+(x-y)^2}\right)^4
=-\frac{i\omega^2\log\t\varepsilon}{2\pi^2\sqrt{1+\omega^2}}\frac{x_f^4}{y^4(x_f-y)^4}\,.
}
Therefore, the final expression for the three-point correlator is
\eq{
\langle{\cal O}_A(0){\cal O}^*_A(x_f){\cal L}(y)\rangle\approx\frac{I_\phi^{(0)}[\bar{X},\bar{s};y]}{x_f^{2\Delta_A}}
-\frac{iJ^2\log\t\varepsilon}{4\pi^3g\sqrt{J^2+4\pi^2g^2}}\frac{x_f^{4-2\Delta_A}}{y^4(x_f-y)^4}\,.
\label{3pointcir}
}
One can see immediately that \eqref{aLAA} holds, provided that $J$ is kept constant.

\subsection{Giant magnon}

One very important for the AdS/CFT correspondence class of string solutions is the so called giant magnon. In this subsection we will consider the simplest solution of this type~\cite{Hofman:2006xt}. We will use a different parametrization of the sphere
\eq{
ds_{S^5}^2=d\theta^2+\sin^2\theta d\varphi^2+\cos^2\theta d\Omega^2_3\,.
}
We start with the ansatz suggested in \cite{Hofman:2006xt}
\eq{
\cos\theta=\sin\frac{p}{2}\,{\rm sech}(\omega u)\,,\quad\tan(\varphi-\omega\tau)=\tan\frac{p}{2}\tanh(\omega u)\,,
\label{mag-ans}
}
where $u=\left(\sigma-\tau\cos\frac{p}{2}\right)\csc\frac{p}{2}$, and $p\in[0,2\pi)$ is the momentum of the magnon. The angular momentum of the string becomes
\eq{
J=2g\int_{-L}^{L}d\sigma\,\sin^2\theta\,\dot{\varphi}=2g\omega\int_{-L}^{L}d\sigma\,\tanh^2(\omega u)\approx 4g\!\left(\omega L-\sin\frac{p}{2}\right)
}
in the limit of large $L$. The energy can be obtained to be
\eq{
E=\Delta^{(0)}=J+4g\sin\frac{p}{2}\approx4g\omega L\,.
}

In this case \eqref{Z2} assumes the form
\eq{
a_2[\bar{X},\bar{s},\Phi]=-\frac{2\omega^2}{\pi}\int_{-\bar{s}/2}^{\bar{s}/2}d\tau\int_{-L}^{L}d\sigma\,e^{\phi/2}\tanh^2(\omega u)\,,
}
where we have used that the saddle point is \cite{Costa:2010}
\eq{
\label{saddlemagnon}
\bar{s}=\frac{2i}{\omega}\log\frac{\varepsilon}{x_f}\,,
}
which again due to \eqref{kappabc} implies the Virasoro constraint $\kappa=i\omega$. The modified correction \eqref{tildea2} takes the form
\eq{
\tilde{a}_2[\bar{X},\bar{s},\Phi=0]=\frac{8i\omega L}{\pi}\log\frac{\varepsilon}{x_f}\,.
}
Therefore we find that the contribution of $\tilde{Z}(0,x_f,\Phi_0=0)$ is
\eq{
\langle{\cal O}_A(0){\cal O}^*_A(x_f)\rangle\sim\left(\frac{\varepsilon}{x_f}\right)^{2\Delta_A}=\left(\frac{\varepsilon}{x_f}\right)^{2\Delta_A^{(0)}-\frac{4i\omega L}{\pi}\log\t\varepsilon}\ \ \Longrightarrow\ \
\Delta_A^{(2)}\approx-\frac{2i}{\pi}\!\left(\frac{J}{4g}+\sin\frac{p}{2}\right)\log\t\varepsilon\,,
}
where we have used that $\Delta_A=\Delta_A^{(0)}+\Delta_A^{(2)}$.

In order to obtain the correction to the three-point correlator, we evaluate \eqref{Iphi} at the saddle point \eqref{saddlemagnon}
\begin{align}\nonumber
I_\phi^{(2)}[\bar{X},\bar{s};y]&=\frac{3\omega^2\log\t\varepsilon}{\pi^3}\int_{-\bar{s}/2}^{\bar{s}/2}d\tau\int_{-L}^{L}d\sigma\,\tanh^2(\omega u)\left(\frac{z}{z^2+(x-y)^2}\right)^4\\
&=\frac{3\omega J\log\t\varepsilon}{2\pi^3g}\int_{-\bar{s}/2}^{\bar{s}/2}d\tau\left(\frac{z}{z^2+(x-y)^2}\right)^4
=-\frac{iJ\log\t\varepsilon}{8\pi^3g}\frac{x_f^4}{y^4(x_f-y)^4}\,.
\end{align}
For the three-point function we end up with the expression
\eq{
\langle{\cal O}_A(0){\cal O}^*_A(x_f){\cal L}(y)\rangle\approx\frac{I_\phi^{(0)}[\bar{X},\bar{s};y]}{x_f^{2\Delta_A}}
-\frac{iJ\log\t\varepsilon}{8\pi^3g}\frac{x_f^{4-2\Delta_A}}{y^4(x_f-y)^4}\,.
}
Again it can be seen that \eqref{aLAA} holds (up to zeroth order of $L$), provided that $p$ and $J$ are kept independent of the coupling.

\sect{Conclusion}

One of the active fields in AdS/CFT correspondence recently has been the calculation of three-point correlators beyond the supergravity approximation \cite{Janik:2010gc,Zarembo:2010,Costa:2010,Roiban:2010}. The authors consider string theory on $\axs$ and obtain three-point functions of two ``heavy'' operators and one supergravity state at strong coupling. Very recently three-point correlators of three BMN operators with large charges were computed \cite{Klose:2011}, albeit in light-cone gauge.

In this paper we generalize the ideas for calculation of correlation functions in \cite{Janik:2010gc,Costa:2010} by including the leading bosonic quadratic correction to the action. We apply the method in the cases of simple string solutions. The careful analysis in \cite{Costa:2010} states that the structure constants are given by $-g^2\partial\Delta_A/\partial g^2$. We are the first to check this relation for quadratic corrections to correlators, and we found perfect agreement.

Unfortunately, to the best of our knowledge, there are no known results from gauge theory side which can be used for comparison, although there has been progress in the computation of structure constants using the Bethe eigenstates of the underlying integrable spin chain in the weak coupling limit of the dual gauge theory \cite{Escobedo:2010,Escobedo:2011}. The corrections to correlation functions we obtained here should be considered only as a first step. For example, it would be interesting to extend our results to other ``light'' operators\footnote{We note, however, that in the cases of other ``light'' operators a different approach should be used, probably involving vertex operators.} and backgrounds, as in \cite{Klose:2011}.

\section*{Acknowledgments}
The authors would like to thank H. Dimov for valuable discussions and careful reading of the paper. This work was supported in part by the Austrian Research Funds FWF P22000 and I192, and NSFB DO 02-257.


\end{document}